# GPT Chatbots for Alleviating Anxiety and Depression: A Pilot Randomized Controlled Trial with Afghan Women

**Sofia Sahab[1], Jawad Haqbeen[1], Diksha Sapkota[2], and Takayuki Ito[1]**

[1]Department of Social Informatics, Kyoto University, Kyoto, 606-8501 Japan

[2]Griffith Criminology Institute, Brisbane, 4113 Australia

Corresponding author: Sofia Sahab (e-mail: sahab.sofia@i.kyoto-u.ac.jp).

This research was supported partially by the JST CREST fund (Grant Number: JPMJCR20D1, Japan) and JSPS KAKENHI (Grant Number: 22K17948, Japan).

**ABSTRACT** In this study, we investigated the effects of GPT-4, with and without specific conversational instructions, on the mental health of Afghan women. These women face multifaceted challenges, including Taliban-imposed restrictions, societal inequalities, and domestic violence, adversely affecting their well-being. We conducted a randomized controlled trial with 60 participants, dividing them into three groups: GPT-4, a supportive listener (GPT-4 with empathetic engagement instructions), and a waiting list. The Hospital Anxiety and Depression Scale (HADS) was used to measure anxiety and depression before and after the intervention. Linguistic analysis of chat data examined personal pronouns, tones, emotions, and Language Style Matching (LSM). The supportive listener group showed a significant reduction in HADS scores compared to the other groups. Linguistic analysis revealed a more positive tone and higher LSM in the supportive listener group, with a significant negative correlation between LSM and changes in HADS scores, indicating greater linguistic alignment was linked to reductions in anxiety and depression. Perceived empathy ratings were also significantly higher in the supportive listener group. These findings highlight the potential of AI-driven interventions, like GPT-4, in providing accessible mental health support. However, such interventions should complement traditional psychotherapy, ensuring a collaborative approach to optimize therapeutic outcomes.

**INDEX TERMS** AI driven interventions, anxiety and depression, GPT-4, mental health, linguistic analysis, psychotherapy.

## I. INTRODUCTION

Mental illness is a growing public health concern worldwide, with 80% of people living with a mental health disorder residing in Low- and Middle-Income Countries (LMICs) [1, 2]. Women experience higher levels of mental health disorders than men, and this gender disparity tends to widen with age [3]. The persistently high prevalence of mental disorders is alarming as it impacts individuals' health and daily lives, while also causing ripple effects across families, communities, and national economies [4]. The economic burden of mental health disorders stems from the high cost associated with healthcare and loss of productivity, with an anticipated toll in LMICs between 2011 and 2030 estimated at US$ 7.3 trillion [5].

Despite the current political constraints in Afghanistan that limit mental health surveys, available studies indicate a significant mental health burden among Afghan women. For example, a recent survey found that 47% of Afghan women experienced high psychological distress [6]. Another cross-sectional study reported that nearly 80% of Afghan women exhibited depressive symptoms [7]. This high prevalence of mental illness among Afghan women is largely attributed to severe trauma, interpersonal violence, poverty, limited opportunities for employment and education, and prevailing patriarchal norms in society [8, 9]. Afghan women have also endured the effects of forty years of war and conflict, and the recent Taliban takeover of the government in August 2021 has further restricted their basic human rights, such as education and work [10, 11]. These factors significantly impact their mental health and overall well-being. However, cultural and societal norms often hinder seeking assistance for mental health issues [12]. Sharing one's problems, even with friends and family members, is often not considered acceptable behavior [12].

Despite increasing efforts to provide mental health services for people in need, most countries lack the necessary human and financial resources, and the situation is particularly worse in LMICs [13]. As of 2020, there are 13 mental health workers per 100,000 population globally, compared to 1.6 per 100,000 in Africa and 2.8 per 100,000 in South Asia [14]. There is also a dire shortage of mental health professionals in Afghanistan; as of 2016, there were 0.231 psychiatrists and 0.296 psychologists working in the mental health sector per 100,000 population [15]. To address the acute shortage of mental health professionals, high burgeoning costs associated with mental health services, and social stigma regarding mental help-seeking, there has been a rapid increase in technology-based treatment options globally [16, 17].

One such advancement is the use of chatbots, also known as conversational agents, to converse and interact with humans [18, 19]. Previous studies assessing the effectiveness of mental health chatbots have reported mixed results. For example, Inkster and colleagues found that Wysa, an empathy-driven conversational artificial intelligence agent, significantly reduced depressive symptoms among users [20]. Similarly, a pilot study in Sweden reported that participants who adhered to the intervention that have used a conversational interface via an automated smartphone-based chatbot showed improved psychological wellbeing and reduced stress compared to a wait-list control group [21]. Conversely, users of ChatPal, a multilingual mental health and well-being chatbot, did not exhibit significant improvements in their well-being scores after a 12-week intervention and reported technical issues [21]. Despite mixed findings on effectiveness, there is consensus across studies that these interventions are generally acceptable to users, with high engagement and retention rates [21, 22].

Although Artificial Intelligence (AI)-powered chatbots have potential mental health benefits, current knowledge on this topic remains preliminary and necessitates further research. Significant variations exist in the interventions used (e.g., content, duration, and target groups) and the outcomes assessed. Moreover, available studies are primarily from high-income countries, limiting the generalizability of findings to low-income countries. Another notable gap in the literature is the lack of studies examining the impact of such interventions on women specifically. Gender and socio-cultural contexts significantly influence how individuals express themselves and seek support. Evidence suggests that women face gendered pathways to mental illness, including a history of childhood maltreatment, domestic and family violence, limited education, underemployment, and parenting responsibilities [23-26].

Existing chatbots focusing on gender-specific issues, such as those for domestic violence survivors, often prioritize providing legal resources or general guidance, such as advice on reporting abuse or finding local services, rather than offering therapeutic interventions aimed at addressing mental health concerns [27, 28]. There is a notable gap in AI interventions specifically designed to address the mental health needs of women experiencing gendered traumas like Domestic and Family Violence (DFV), particularly when compounded by systemic inequalities. This study aims to fill this gap by designing and evaluating an AI-powered chatbot intervention, employing a randomized controlled trial design, and tailoring it to Afghan women, a highly underserved population uniquely positioned to benefit from accessible, scalable mental health solutions.

Two types of chatbots were assessed: a supportive listener chatbot and a standard Generative Pre-trained Transformer-4 (GPT-4) chatbot. The supportive listener chatbot, based on GPT-4, was specifically instructed to maintain a non-judgmental approach, avoid invasive or pushy questions, and refrain from prompting users to recall traumatic memories. It used simple and empathetic language suitable for non-native English speakers, creating a safe and supportive environment for psychological support.

In contrast, the standard GPT-4 chatbot operated without specialized instructions for empathetic interactions. Both chatbots did not retain conversation memory within sessions to ensure consistent performance, respect user privacy, and mitigate potential mental pressure from sensitive topics. Additionally, a waiting list group was included as a control to compare the effectiveness of the two chatbots.

Our study was designed to investigate the following hypotheses:

H1: Utilizing GPT-4 with simple instructions as a supportive listener leads to a significant decrease in Hospital Anxiety and Depression Scale (HADS) scores among Afghan women experiencing challenges such as Taliban-imposed restrictions, societal inequalities, and domestic violence.

H2: Utilizing GPT-4 with simple instructions for empathetic engagement, but without explicit guidance on language aspects such as tone or emotion, yields a more positive tone and emotion in its interactions with participants compared to GPT-4 interactions without such instructions.

H3: Higher Language Style Matching (LSM) scores in Human-chatbot dyads associates with a reduction in anxiety and depression in participants post-intervention compared to pre-intervention.

We also explored perceived empathy in chatbot interactions, assessing whether participants perceived differences in empathy between the two chatbots.

## II. RELATED WORK

The use of technology for mental health and wellbeing is becoming increasingly common, driven by its widespread availability, scalability, and affordability [16, 29]. From online platforms offering therapy sessions conducted by psychiatrists to more sophisticated approaches integrating human and technological elements, such as AI-based support systems, a range of solutions have been developed to meet the diverse needs of individuals seeking mental health support [16, 17, 30]. Among these advancements, there are specific technologies tailored for survivors of domestic violence and women who have endured various forms of trauma [31, 32]. In this section, we review some of the most prominent technologies in this domain and identify the gap that our study aims to address.

The initial wave of platforms leveraging technology for mental health support primarily relies on human interaction. These platforms offer therapy sessions through video calls, chat sessions, and digital resources, delivered by licensed professionals [33]. This approach allows individuals to access therapy from remote locations, potentially making it more affordable by eliminating transportation costs [34, 35]. However, despite these advantages, the human-centric nature of these platforms presents limitations. The availability of therapists may still be insufficient to meet the needs of large populations, and while eliminating transportation costs, the fees associated with human therapists may remain prohibitive for many [17]. Stigma associated with mental illness is another significant barrier to seeking help from therapists [17, 36, 37].

Considering these challenges, a new wave of platforms supported by AI tools has emerged [20, 37]. AI-supported platforms are being proposed as a viable solution to enhance scalability to reach a wide audience, affordability, immediate availability, and reduce societal stigma in the provision of mental health care [16, 17, 30]. These interventions leverage machine learning algorithms and natural language processing to simulate human-like interactions and provide support to individuals [38]. While Cognitive Behavioral Therapy (CBT) is predominantly utilized by popular chatbots such as Woebot, Wysa, Youper, and Tess, a variety of therapeutic approaches including positive psychology, Dialectical Behavior Therapy (DBT), motivational interviewing, positive behavior support, and behavioral reinforcement are also incorporated into these AI-supported platforms [17, 30].

Although chatbots like Woebot and Wysa, developed for general therapy, can broadly address mental health issues arising from gender-based violence and other forms of trauma experienced by women, there are chatbots specifically designed for this purpose, however they are only a few [39]. In these specialized chatbots, therapy is often of less importance compared to other critical aspects such as submitting evidence of abuse or answering Domestic Violence (DV)-related questions [27, 28]. They are generally designed to provide survivors with an environment to get help, learn about their rights, and find resources like nearby shelters and hotlines [40]. For example, #MeTooMaastricht provides survivors of sexual harassment with necessary information and directs them to appropriate institutions [27]. Similarly, the Conversational Interactive Response (CIR) in the mobile application developed by Hossain et al. for DV survivors displays questions and provides instant responses, such as the number of a nearby police station or advice on what to do during an episode of violence [28]. This gap underscores the need for AI-powered interventions that provide sustained, culturally sensitive, and therapeutic mental health support tailored to the unique challenges faced by women, particularly those in underserved and high-stigma contexts.

The effectiveness of such AI-powered interventions depends significantly on the underlying technical frameworks employed in chatbot development. Two primary approaches are employed in the development of chatbots: rule-based and Machine Learning (ML)-based [41, 42]. Rule-based chatbots rely on predefined rules and scripts to simulate conversations [30, 41, 42] . They follow a tree-based structure where each user input triggers a specific response based on pre-written dialogue [30]. A recent study by Grodniewicz and Hohol (2024) indicates that current therapeutic chatbots, such as Woebot, Wysa, Youper, and Tess, operate within this rule-based framework [30]. It can be assumed that their conversational pathways are dictated by fixed rules, adhering to a structured flowchart design [30]. These chatbots are typically limited to the scenarios and responses they were programmed with, making them less flexible but easier to control [41]. On one side, it is advantageous for controlling user input and AI output; on the other side, limiting users to only a few options may not be conducive to a natural and engaging conversation [30, 41, 42].

ML-based chatbots offer greater flexibility and adaptability compared to rule-based chatbots [41]. These chatbots use algorithms to learn from data and improve over time, allowing them to handle a wider range of user inputs and provide more nuanced responses [42]. ML-based chatbots can understand the context of a conversation, grasp the intent behind a user's query, and generate more human-like responses [42]. This approach enables them to handle variations in language and user queries, making interactions with users more natural and effective [41]. However, ML-based chatbots require a large amount of training data and continuous refinement to maintain their performance and accuracy [30, 41].

ChatGPT, developed by OpenAI, exemplifies this technology, utilizing machine learning algorithms and trained on extensive text datasets with billions of parameters to model predictions and decisions [30]. Large Language Models (LLMs) like ChatGPT, GPT-4, and BERT (Bidirectional Encoder Representations from Transformers ) are designed to process and generate text across a wide range of topics [30]. As such, they reach or surpass human psychologists in terms of social intelligence [43]. Given these advantages, ChatGPT has the potential to significantly enhance the exchange of information between psychiatry and AI [44]. Yet, the current applications of ChatGPT in psychiatry primarily revolve around aiding psychiatrists with their everyday tasks [44]. Platforms such as Kokobot and ChatBeacon are working to incorporate this advanced AI into mental health support systems [45]. However, comprehensive evaluations of its effectiveness through randomized controlled trials are currently lacking. This study addresses these critical gaps by designing and evaluating an AI-powered chatbot intervention tailored to the mental health needs of Afghan women. The chatbot utilizes GPT-4 customized with instructions to emphasize empathetic communication, avoid intrusive questioning, and foster a supportive environment. By doing so, this study aims to explore the chatbot's potential to alleviate anxiety and depression among vulnerable populations. In addition, we investigate the emotional tones and Linguistic Style Matching (LSM) between users and the chatbots.

## III. METHODS

### A. ELIGIBILITY CRITERIA AND STUDY PARTICIPANTS

Participants needed to meet the following criteria for inclusion in the study: be Afghan residents aged 18 years or older, female, with a literacy level above high school, owning an internet-enabled mobile phone or computer, possessing a valid email address and phone number, proficient in English, and reporting scores indicating moderate to poor well-being in the 5-item World Health Organization Well-Being Index (WHO-5; total score ≤ 19) and moderate to low self-efficacy (total score ≤30).

Participants were recruited online through a respondent recruiting agency in Afghanistan. The agency announced the call for participation on their online job portal, specifying the need for female participants only. The registration process included a consent form, demographic questions, and inquiries about English proficiency, as the AI chat required participants to be proficient in English, along with questions related to general self-efficacy and WHO-5 wellbeing.

A total of 2,293 participants registered between December 25, 2023, and January 3, 2024. Of these, 2,176 participants were excluded: 513 due to incomplete responses, 52 for registering multiple times, 27 for not residing in Afghanistan, 736 for being male, 184 for lacking English proficiency, 185 for having a literacy level below high school, and 479 for having high scores on the WHO-5 and self-efficacy assessments, exceeding the thresholds specified in the inclusion criteria. After applying these criteria, 117 participants remained, from which 60 participants were randomly selected and assigned to the control and treatment groups (see Table I for demographics of participants) for the purpose of the current study. Randomization was conducted in Excel using the RAND() function followed by sorting. Among the selected participants, 40 participants were also included in another study [46], with 20 participants unique to each study.

Participants received US$20/AFN 1,400 as compensation for their participation in the chat with the chatbot and for completing surveys.

### B. STUDY DESIGN

The study employed a Randomized Controlled Trial (RCT) design to investigate the effects of chatbot interactions on anxiety and depression among women. Participants were randomly assigned to one of three conditions: chatting with a GPT-4 chatbot, chatting with a supportive listener chatbot, or being placed on a waiting list (control group).

TABLE I
SAMPLE DEMOGRAPHICS

| Category | Number | Percentage |
|---|---|---|
| Age | | |
| 18-22 | 15 | 25.00 |
| 23-27 | 35 | 58.33 |
| 28-32 | 8 | 13.33 |
| 33-37 | 2 | 3.33 |
| Education | | |
| Bachelor's degree holder | 19 | 31.67 |
| Undergraduate student | 38 | 63.33 |
| graduate student | 3 | 5.00 |

### C. MEASURES

Symptoms of anxiety and depression were assessed using the 14-item Hospital Anxiety and Depression Scale (HADS) [47]. This scale comprises seven items designed to gauge depression symptoms and seven to evaluate anxiety symptoms. Participants self-reported their symptoms using a four-point Likert scale, with responses ranging from 0 to 3. The HADS measurements were administered at two time points for each participant: before the intervention (Time 1) and one week after the intervention (Time 2).

To select subjects experiencing some levels of anxiety and depression, considering that the HADS were too detailed for the registration phase, we used the WHO-5 and General Self-Efficacy Scales during registration. The 5-item World Health Organization Well-Being Index (WHO-5) 1998 version is a concise and universal measure for assessing subjective well-being. Each question is rated from 0 to 5. The raw score is calculated by totaling the figures of the five answers, ranging from 0 to 25, with 0 representing the worst possible and 25 representing the best possible quality of life.

Participants' self-efficacy was assessed using the Generalized Self-Efficacy Scale [48], which comprises 10 items. Responses were recorded on a four-point Likert scale, ranging from 1 to 4. The final composite score, with a range from 10 to 40, is derived by summing up the responses to all 10 items.

Artificial empathy was measured using the RoPE (Robot's Perceived Empathy) scale [50], which was originally developed for robots and later adapted for artificial agents [51]. The scale consists of two subscales: Empathic Understanding (8 items) and Empathic Response (8 items), along with 4 filler items. Participants responded using a Likert scale ranging from -3 (No, I strongly feel that it is not true) to 3 (Yes, I strongly feel that it is true). The total score is calculated by summing the values of each item after reversing the negatively coded items.

### D. CHATBOT INTERACTION TOOL

To enable participants to chat with the GPT-4 or the supportive listener chatbots, we developed both front-end and back-end components, in addition to setting up a cloud server infrastructure. The front-end, created using JavaScript, included a chatbot interface for displaying chatbot messages and a user dialogue box. This interface was also preloaded with specific preliminary information, crucial for assessing the user's status and safety before initiating a dialogue.

The back-end featured the GPT-4 and the supportive listener agents, which were interfaced with OpenAI's Assistant Application Programming Interface (API). While the GPT-4 functioned with its standard capabilities without specific instructions, the supportive listener, leveraging the Assistant API, was programmed to prioritize attentive listening and empathetic engagement with the user's emotions. The supportive listener was instructed to provide psychological support, avoiding repetitive questioning and refraining from prompting survivors to recall traumatic memories. It was also instructed to avoid asking personal, invasive, or pushy questions, respecting the privacy and boundaries of the participants. It was to adopt a non-judgmental approach, respecting the choices of participants without commenting on their decisions, even if they may seem unwise. It was to use language that is simple and empathetic, suitable for non-native English speakers, fostering an informal and supportive conversation aimed at providing a safe environment. These instructions were designed to align with the needs of Afghan women and were informed by research on conversational agents for sensitive populations, such as survivors of domestic violence [40].

For example, when participants shared that they were feeling depressed, the supportive listener responded empathetically as follows:

> *I'm really sorry to hear that you're feeling this way. It's not uncommon to experience periods where things can feel overwhelming. Remember, it's okay to not be okay, and reaching out as you're doing now is a courageous step. While I'm here to listen and support you, I highly recommend you consider reaching out to a qualified professional who can provide you with the help and support you need to navigate this challenging time. Depression is a serious condition, but with the right support, it can be managed. In the meantime, I'm here to keep you company. If you want to talk about what you're feeling, or if there are other subjects you'd like to discuss or questions you have, please feel free to share. You don't have to go through this alone.*

The supportive listener’s response provides emotional support while gently recommending professional help. This approach is particularly crucial given the stigma surrounding mental health in Afghanistan, where direct suggestions to seek professional help may not always be well-received [49]. Furthermore, the response’s indirect and gentle tone aligns with Afghanistan’s high-context communication culture, which values subtlety and emotional sensitivity [50].

In contrast, GPT-4, operating without specific instructions, responded with:

> *I'm really sorry that you're feeling this way, but I'm unable to provide the help that you need. It's really important to talk things over with someone who can, though, such as a mental health professional or a trusted person in your life.*

This response, while acknowledging the issue, is more direct and does not engage in active listening before recommending external help. Without first creating a supportive context, it risks discouraging individuals from opening up further, especially in a cultural setting like Afghanistan, where seeking help is stigmatized [49].

To further ensure cultural and linguistic appropriateness for Afghan women, the chatbot's responses were reviewed and tested by two native Afghan authors to ensure relevance, sensitivity, and an appropriate conversational tone.

Both chatbots did not retain conversation memory within sessions to ensure consistent performance, respect user privacy, and mitigate potential mental pressure from sensitive topics. The back-end was implemented using Python, and the deployment of the system's front-end and back-end was executed within an Elastic Compute Cloud (EC2) instance on Amazon Web Services (AWS).

### *E. PROCEDURE*

Participants were randomly assigned to three groups: chat with GPT-4, chat with supportive listener, and a waiting list. The recruiting company contacted participants and provided information about the experiment's procedures, compensation, and confidentiality. Participants were also given instructions on how to chat with the chatbot and fill out questionnaires and were provided with their assigned IDs and group IDs.

At the beginning of the study, participants completed a questionnaire that included a consent form, demographic questions, and HADS. The following day, the chat groups interacted with their respective chatbots for one hour, after which they completed an empathy assessment. A follow-up assessment was conducted one week after the chat, during which participants completed the same measures along with additional filler questions. The waiting list participants engaged with the supportive listener on the subsequent day. The experiment took place from January 9 to 18, 2024.

Chat with the chatbot was conducted in four batches on the same day, with each batch consisting of 10 participants (five for each chat group) simultaneously. This was due to system limitations not optimized for simultaneous access by many users. The chat link was shared with participants 10 minutes before the session. The chat started with the chatbot introducing itself, inquiring about participants' safety, asking for their IDs and groups, and obtaining data usage consent. Participants were then asked to share any challenges, concerns, or problems they would like to discuss. After the one-hour chat, participants were instructed to end the chat and close the webpage.

Real-time monitoring of conversation logs ensured a minimum chat duration of 50 minutes.

### F. DATA ANALYSIS

Repeated measures Analysis of Variance (ANOVA) was selected as the appropriate statistical test to analyze differences in the dependent variable (HADS), with a between-subject factor (condition) and a within-subject factor (time). A significance level of 0.05 and two-tailed tests were used for all statistical analyses.

The statistical analysis was conducted using IBM SPSSStatistics version 29.0.1.0. For linguistic analysis, we used Linguistic Inquiry and Word Count (LIWC-22) version 1.5.0.

For the word cloud analysis, we excluded several frequently used words that did not contribute to a clear understanding of the shared problems. These words included "thank," "thanks," "Afghanistan," "problems," "problem," "situation," "find," "girl," "girls," "woman," "women," "time," and "solution." Words such as "problem" or "thank" were excluded because they were overly generic and frequently used in contexts that did not offer meaningful insights into the specific challenges or emotions expressed by participants. Others, such as "Afghanistan" and "girl" or "woman," were omitted due to their limited semantic contribution. For instance, "Afghanistan" often appeared as a geographic reference, contextualizing the societal and cultural environment in which these challenges occurred, rather than representing a specific issue itself. Likewise, "girl" and "woman" referred broadly to the study's target demographic, drawing attention to the unique difficulties faced by women in Afghanistan without providing deeper insights into those challenges. This filtering approach ensured that the analysis focused on words that provided greater specificity and depth to the participants' narratives. In addition to these words, the stop list used for this analysis incorporated function words (e.g., articles, prepositions, pronouns) from the LIWC default list, which are common in language but do not contribute substantively to the overall understanding of the text.

To compare language use between the control and treatment groups, we analyzed the chat data of the chatbots interacting with the participants. This analysis focused on pronouns ('I,' 'we,' and 'you'), tone categories (positive and negative), emotion categories (positive, negative, anxiety, anger, and sadness), and Language Style Matching (LSM). LSM measures the similarity in the usage frequency of function words between participants and the chatbot, providing insights into linguistic alignment during interactions.

### G. ETHICAL CONSIDERATIONS AND APPROVAL

Due to the non-recognition of the existing government in Afghanistan by the international community and the limitations they have imposed on women, seeking approval from a local ethics review committee in Afghanistan was not feasible. Therefore, our study received ethics approval from the Ethics Committee of the Graduate School of Informatics, Kyoto University (KUIS-EAR-2023-004).

Informed consent was obtained from all participants. Prior to commencing any research activities, participants received comprehensive information about the study's procedures, including surveys and chatting with a chatbot, as well as details of the research team and the affiliated institution. They were also informed about the study's primary goal of providing support to Afghan women by listening to their problems. No deception was involved; participants were aware that they would be interacting with a chatbot before chatting with the agent. At the outset of each interaction, the chatbot explicitly disclosed its nature as an AI chatbot and not a human. The chatbot was named "ChatBot".

## IV. RESULTS

### A. HOSPITAL ANXIETY AND DEPRESSION SCALE (HADS) OUTCOMES

This section presents the results from the repeated measures ANOVA conducted on the Hospital Anxiety and Depression Scale (HADS), including its subscales for Anxiety (HADS-A) and Depression (HADS-D), as well as the overall HADS score.

#### 1) ANXIETY OUTCOMES (HADS-A)

The repeated measures ANOVA for HADS-A revealed a significant 3 (groups) by 2 (times) interaction effect, Wilks' $\Lambda = 0.90$, $F(2, 57) = 3.19$, $p = 0.048$, $\eta_p^2 = 0.10$. This indicates that anxiety scores changed significantly over time between the different intervention groups (see Table II).

#### 2) DEPRESSION OUTCOMES (HADS-D)

The repeated measures ANOVA for HADS-D demonstrated a significant 3 (groups) by 2 (times) interaction effect, Wilks' $\Lambda = 0.88$, $F(2, 57) = 3.71$, $p = 0.031$, $\eta_p^2 = 0.12$. This suggests significant changes in depression scores over time among the groups (see Table II).

#### 3) OVERALL HADS OUTCOMES

The repeated measures ANOVA for overall HADS scores indicated a significant 3 (groups) by 2 (times) interaction effect, Wilks' $\Lambda = 0.83$, $F(2, 57) = 5.91$, $p = 0.005$, $\eta_p^2 = 0.17$.

Post hoc Bonferroni-adjusted pairwise comparisons were conducted to further explore the significant interaction. These comparisons showed a significant decrease in HADS scores for the supportive listener group (mean difference between pre- and post-intervention = 2.60, 95% *CI* [0.14, 5.06], $p = 0.038$, Cohen's $d = 0.47$). In contrast, the GPT-4 group exhibited a significant increase in HADS scores (mean difference = -3.15, 95% *CI* [-5.61, -0.69], $p = 0.013$, Cohen's $d = -0.57$). However, there was no evidence of a statistically significant difference in HADS scores between pre- and post-intervention for the waiting list group (mean difference = -1.65, 95% *CI* [-4.11, 0.81], $p = 0.184$, Cohen's $d = -0.30$) (see Table III).

TABLE II
ANOVA RESULTS FOR HADS-A, HADS-D, AND OVERALL SCORES

| Measure | Wilks' $\Lambda$ | $F$-value | $p$-value | $\eta_p^2$ |
|---|---|---|---|---|
| HADS-A (Anxiety) | 0.90 | 3.19 | 0.048 | 0.10 |
| HADS-D (Depression) | 0.88 | 3.71 | 0.031 | 0.12 |
| HADS Overall | 0.83 | 5.91 | 0.005 | 0.17 |

TABLE III

CHANGES IN HADS SCORES ACROSS GROUPS BEFORE AND AFTER THE INTERVENTION ($N$=20 PER GROUP)

| Group | Mean Score (Pre) (SD) | Mean Score (Post) (SD) | Mean Difference (Pre-Post) | 95% Confidence Interval for Difference[a] | *p*-value[a] | | Cohen's *d* |
|---|---|---|---|---|---|---|---|
| Supportive Listener | 19.65 (5.15) | 17.05 (5.12) | 2.60 | [0.14, 5.06] | 0.038 | * | 0.47 |
| GPT | 17.35 (6.35) | 20.50 (4.95) | -3.15 | [-5.61, -0.69] | 0.013 | * | -0.57 |
| Waiting List | 18.55 (5.99) | 20.20 (5.28) | -1.65 | [-4.11, 0.81] | 0.184 | | -0.30 |

Notes: [a] Adjustments for multiple comparisons are made using the Bonferroni method, * $p < 0.05$.

## B. LINGUISTIC ANALYSIS OUTCOMES

To gain an understanding of the issues shared by participants, a word cloud was generated (see Fig. 1). This cloud provides a general image of the problems that the participants discussed with the agent. The most frequently used words, such as "work" (38), "education" (37), "life" (35), "Taliban" (33), "job" (29), "study" (26), "family" (23), "stress" (21), "current" (21), and "sleep" (20), highlight the key themes that emerged during the chat sessions.

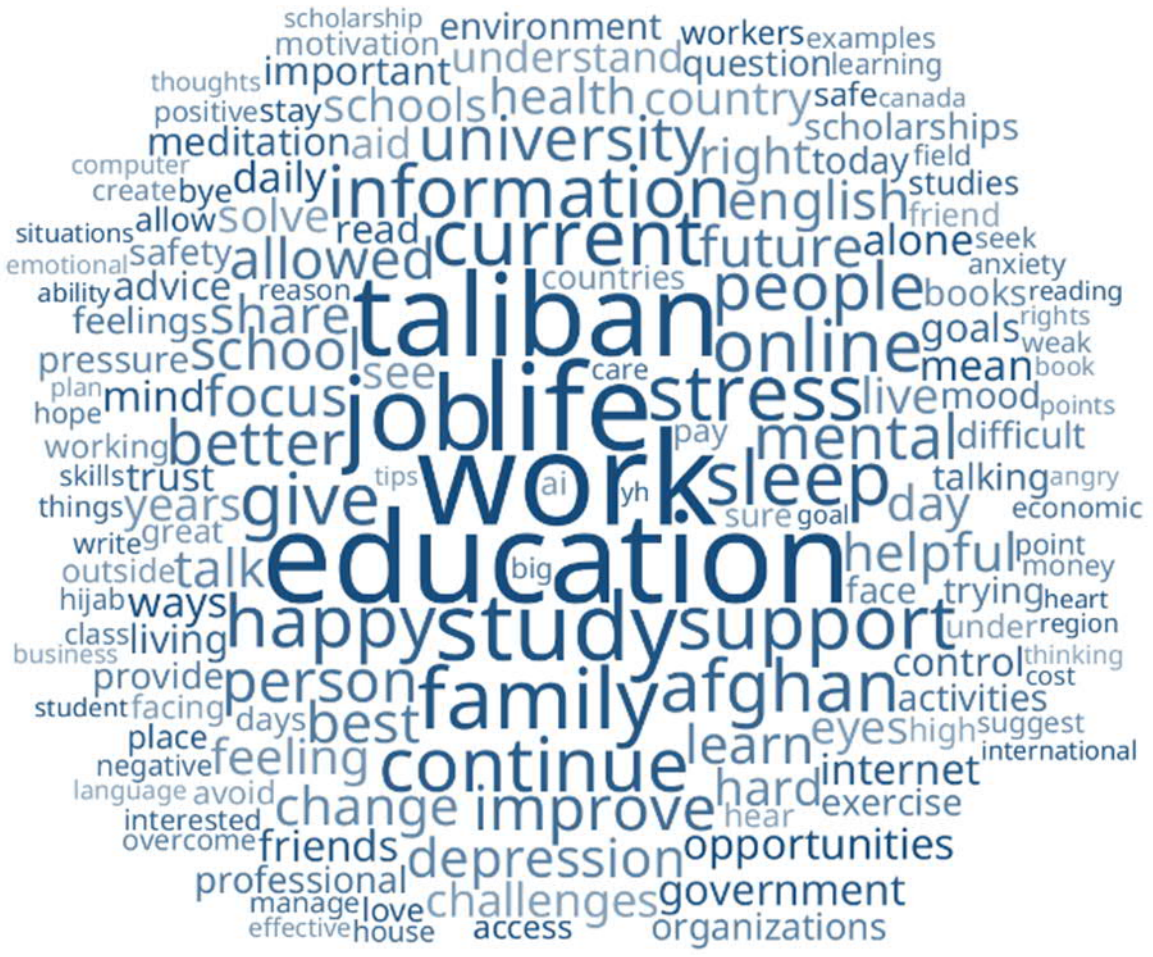

**FIGURE 1.** **Word cloud depicting the most frequently used words by participants in chat sessions with the chatbot. The size of each word corresponds to its frequency of use, providing a visual representation of key themes discussed during the interactions.**

Following the word cloud analysis, we conducted a detailed examination of the chat data to explore language use differences. This analysis focused on the frequency of pronouns ("I," "we," and "you"), tone categories (positive and negative), and emotion categories (positive, negative, anxiety, anger, and sadness) in the chat data of the chatbots and human participants.

When comparing the two chatbots' linguistic interactions, significant differences were found. Specifically, the analysis of the chatbot chats showed that the supportive listener used the pronoun "we" less frequently compared to the GPT-4 ($t(38) = 2.27$, $p = 0.029$, Cohen's $d = 0.72$), and used the pronoun "you" more frequently ($t(38) = -2.61$, $p = 0.013$, Cohen's $d = -0.83$).

Furthermore, while there was no significant difference in the frequency of negative tone and emotion between the two chatbots ($t(29) = 0.49$, $p = 0.626$, Cohen's $d$= 0.16 for tone and $t(27) = 0.92$, $p = 0.364$, Cohen's $d$= 0.29 for emotion), the supportive listener exhibited a higher frequency of positive tone ($t(38) = -5.14$, $p < 0.001$, Cohen's $d = -1.63$) and positive emotion ($t(38) = -2.29$, $p = 0.028$, Cohen's $d = -0.72$) compared to the GPT. The statistical results and summary metrics for the analyzed variables are presented in Table IV.

We found no significant differences between the participants' interactions in the GPT-4 and supportive listener groups. This suggests that while participants engaged in similar conversations with regard to content, emotions, and tones, the key differentiator lay in the chatbot interactions.

TABLE IV

LINGUISTIC CHARACTERISTICS OF GPT-4 AND SUPPORTIVE LISTENER CONVERSATIONS (N=20 PER GROUP)

| Linguistic Features | GPT Mean (*SD*) | SL Mean (*SD*) | *t*-value | *p*-value | | Cohen's *d* |
|---|---|---|---|---|---|---|
| **Pronouns** | | | | | | |
| I | 0.56 (0.58) | 0.72 (0.66) | -0.835 | 0.409 | | -0.264 |
| we | 0.20 (0.16) | 0.10 (0.12) | 2.267 | 0.029 | * | 0.717 |
| you | 4.36 (1.68) | 5.78 (1.77) | -2.613 | 0.013 | * | -0.826 |
| **Tone** | | | | | | |
| positive | 5.23 (1.09) | 7.18 (1.30) | -5.143 | < 0.001 | *** | -1.626 |
| negative | 1.62 (0.97) | 1.49 (0.53) | 0.492 | 0.626 | | 0.156 |
| **Emotion** | | | | | | |
| positive | 0.87 (0.45) | 1.23 (0.53) | -2.288 | 0.028 | * | -0.724 |
| negative | 0.87 (0.64) | 0.73 (0.29) | 0.923 | 0.362 | | 0.292 |
| anxiety | 0.48 (0.41) | 0.43 (0.24) | 0.488 | 0.629 | | 0.154 |
| anger | 0.07 (0.11) | 0.03 (0.03) | 1.614 | 0.115 | | 0.510 |
| sadness | 0.12 (0.13) | 0.17 (0.16) | -1.062 | 0.295 | | -0.336 |
| **LSM** | 0.69 (0.08) | 0.75 (0.09) | -2.258 | 0.030 | * | -0.714 |

Notes: SL stands for Supportive Listener and LSM stands for Language Style Matching, * $p < 0.05$, ** $p < 0.01$, *** $p < 0.001$.

## C. LINGUISTIC ALIGNMENT AND ITS IMPACT ON MENTAL HEALTH OUTCOMES

LSM scores were significantly higher in the supportive listener group compared to the GPT-4 group, indicating stronger linguistic alignment, $t(38) = -2.26$, $p = 0.03$, Cohen's $d = -0.71$ (Table 4). Moreover, a statistically significant negative correlation was observed between LSM scores and differences in Hospital Anxiety and Depression (HAD) scores ($p = 0.026$, $r = -0.35$), suggesting that higher LSM scores were associated with reduced HADS score differences.

## D. ARTIFICIAL EMPATHY OUTCOMES

Empathy ratings were higher in the supportive listener condition compared to the standard GPT-4 condition for the Empathic Response subscale. Participants in the supportive listener group rated the chatbot's empathic responses higher than those in the GPT-4 group ($t(38) = -2.401$, $p = 0.021$, Cohen's $d = -0.76$). This suggests that the supportive listener appeared more emotionally responsive, particularly in providing comfort, encouragement, and praise [51].

For the Empathic Understanding subscale, while empathy ratings were slightly higher for the supportive listener chatbot, the difference was not statistically significant ($t(38) = -0.043$, $p = 0.966$, Cohen's $d = -0.01$). This indicates that both chatbots were perceived similarly in terms of empathic understanding.

TABLE V
EMPATHY COMPARISON BETWEEN GPT AND SL GROUPS.

| Scale | GPT Mean (*SD*) | SL Mean (*SD*) | *t*-value | *p*-value | Cohen's *d* |
|---|---|---|---|---|---|
| Empathic Understanding | 12.90 (6.50) | 13.00 (8.17) | -0.043 | 0.966 | -0.01 |
| Empathic Response | 7.40 (5.21) | 12.00 (6.81) | -2.401 | 0.021 * | -0.76 |

Notes: * $p < 0.05$.

## V. DISCUSSION

This study addresses a significant gap in the literature by investigating the use of AI-powered chatbots as a therapeutic intervention tailored specifically for Afghan women, a population facing substantial challenges, including widespread exposure to violence and systemic inequalities. [16, 17, 30]. The use of AI-driven support in a context where traditional mental health resources are limited offers valuable insights into the feasibility and effectiveness of such interventions in low-resource settings [7]. The application of AI for mental health support in Afghanistan highlights its potential for scalability and accessibility, addressing critical needs in a region facing significant challenges [6].

We used GPT-4 with simple instructions as a supportive listener to Afghan women facing challenges such as Taliban-imposed restrictions on education and work, societal inequalities, and domestic violence, all of which could impact their mental health and well-being. The chatbot was instructed to engage in empathetic conversations using simple, non-invasive language, akin to a supportive friend, and avoid triggering traumatic memories by steering clear of pushy or judgmental questions, thus respecting the participants' emotional boundaries and considering their sensitive and intersectional challenges [40].

Our analysis, utilizing HADS, indicated a significant decrease in anxiety and depression scores for the supportive listener group compared to both the GPT-4 group and the waiting list group between pre- and post-intervention, supporting our first hypothesis. There was a significant increase in anxiety and depression scores in GPT-4 group while the waiting list group did not exhibit a statistically significant difference in HADS scores between pre- and post-intervention. This increase in the HADS score in the GPT-4 group may be attributed to the significantly lower perceived empathy, particularly in terms of empathic response, compared to the supportive listener condition [52]. The lack of specific instructions to guide empathetic responses in the GPT-4 interactions likely resulted in a deficit of contextual sensitivity, essential for therapeutic engagement, which could have contributed to heightened anxiety and depression scores. Additionally, linguistic analysis, discussed later, reveals differences in tone, emotional expression, and linguistic alignment, which may further explain these outcomes.

While the intervention yielded positive short-term outcomes (1-week follow-up), we speculate that the effects could potentially be sustained in the long term, similar to findings from Single Session Therapy (SST). Research on SST has shown that brief interventions can have lasting effects, with participants maintaining improvements in anxiety and depression for up to three months post-intervention [53, 54]. However, as discussed in the future directions section, given the scalability and accessibility of AI-driven interventions, more frequent sessions may be necessary to ensure sustained benefits and achieve continuous long-term improvements in mental health.

The calculated effect sizes provide additional clarity on the significance of these findings. The effect size of partial eta squared ($\eta_p^2$) of 0.172 for HADS falls into the range typically considered a large effect size [55]. This is comparable to the effect sizes of other established AI-based mental health interventions, such as Woebot [37] (Cohen's $d = 0.44$) and Wysa [20] (Cohen's $d = 0.47$), which are considered medium effect sizes. Although these studies used the Patient Health Questionnaire-9 (PHQ-9) to measure depression and computed effect sizes for between-group effects, the comparisons are still insightful. Aligning our statistical approach with the Wysa study [20], and conducting a between-groups comparison of the average improvement (HADS at Time 1 minus HADS at Time 2) using an independent samples t-test, yields a Cohen's d of 0.739 for the supportive listener group versus the waiting list group, and a Cohen's $d$ of -1.104 for the supportive listener group versus the GPT-4 group, both of which indicate substantial effects.

The linguistic analysis of chat data provided critical insights and further elucidated these differences. The word cloud analysis highlighted prevalent themes such as concerns about education, work, societal constraints imposed by the Taliban, and familial stressors. These topics underscore the multifaceted challenges faced by Afghan women, shaping the context in which AI interventions operate [56].

Tones and emotions in text-based chatbots are crucial, especially since they lack nonverbal cues like facial expressions or voice tones, which are vital in therapy [57]. Our findings demonstrated that the supportive listener exhibited a more positive tone and emotion compared to GPT-4 alone, despite no specific instructions to do so. This is in line with our second hypothesis and previous research, which has shown that an empathetic tone can alleviate users' negative emotions such as anxiety and sadness and decrease stress [58, 59].

Our analysis of Language Style Matching (LSM), which measures the similarity in the usage frequency of function words [60], revealed significantly higher LSM scores for Human-AI dyads in the supportive listener group (Mean = 0.75) compared to the GPT-group (Mean = 0.69) supporting our third hypothesis. The benchmarks for low and high LSM for human dyads are 0.6 and 0.85, respectively, for daily conversations [61, 62]. While LSM studies on psychotherapy are limited [61], Borelli et al. demonstrated a mean range of 0.87 to 0.89 in psychotherapy sessions [63]. We were unable to locate any research that provided an LSM range for human-AI dyads. However, one study found that AI agents were considered more trustworthy when their conversational style matched that of users with high consideration conversational styles [64]. The observed LSM values for the supportive listener group could serve as a preliminary reference for future investigations into Human-AI dyads, particularly in contexts involving non-native English speakers. The non-native fluency of participants may have influenced their use of function words, which could explain the comparatively lower LSM values relative to those typically seen in human-human psychotherapy settings. Further research is needed to establish standardized benchmarks for LSM in human-AI interactions and to investigate the role of linguistic alignment in enhancing user engagement and improving therapeutic outcomes.

Furthermore, our correlation analysis between LSM scores and changes in HADS scores revealed a significant negative correlation, suggesting that greater linguistic alignment between participants and the AI corresponded with substantial reductions in anxiety and depression scores. This finding is consistent with Borelli et al.'s study, which showed that high early LSM in psychotherapy sessions yielded better outcomes in reducing psychological distress [63], underscoring the importance of linguistic congruence in enhancing therapeutic outcomes within AI-driven interventions. However, further research is needed to determine the causal nature of this relationship and its implications for mental health interventions.

Our study provides significant implications for the advancement of AI-driven interventions using large language models such as GPT in mental health support, particularly for marginalized populations like Afghan women facing multifaceted challenges. The observed effectiveness of GPT-4 as a supportive listener in reducing anxiety and depression underscores the potential for AI technologies to offer accessible and culturally sensitive mental health assistance where traditional resources may be scarce or stigmatized. The findings also highlight the importance of designing AI systems that prioritize linguistic and emotional congruence with users, suggesting avenues for future research to enhance AI's adaptive capabilities in diverse cultural contexts.

### *A. LIMITATIONS AND FUTURE DIRECTIONS*

Despite notable strengths, there are some limitations that warrant consideration while interpreting the findings of the study. As a pilot or feasibility study, the sample size was relatively small, with 20 participants in each group (60 in total), which may limit the generalizability of our findings. While pilot studies are invaluable for testing feasibility and refining intervention protocols, the results cannot be generalized to a larger population. Future research with a larger sample size would be needed to validate our findings and ensure their broader applicability. Additionally, subsequent studies could incorporate a more diverse demographic of Afghan women, such as those from rural versus urban areas or with varying educational backgrounds, to obtain a more representative sample. Moreover, generalizing these results to other cultural contexts requires caution, as mental health interventions can vary significantly across different cultural, social, and economic environments [65]. Future research should also explore the challenges of implementing AI-driven interventions in various settings, considering factors such as language, cultural attitudes toward mental health, and accessibility.

The intervention was of short duration, consisting of a single one-hour session with the AI chatbot. Evaluating the outcomes one-week post-intervention provides only a snapshot of short-term effects, potentially overlooking longer-term impacts that could emerge with sustained engagement. Research on Single Session Interventions (SSIs) suggests that while initial benefits are often significant, these effects tend to diminish over time, particularly beyond 12 weeks [54]. To address this challenge, strategies such as integrating SSIs into adaptive, just-in-time systems that deliver support precisely when users need it, as suggested by Schleider et al. [66], could help sustain the benefits of such interventions. Given the scalability of AI-driven interventions like ours, offering users ongoing or periodic access to sessions may also enhance the longevity of positive outcomes. Future research should explore these approaches to optimize the durability and sustained impact of AI-based mental health supports. Future studies should also incorporate follow-up assessments at multiple time points (e.g., 3 months, 6 months, and 12 months post-intervention) to evaluate sustained effects. These follow-ups could utilize standardized psychological measures such as HADS [47] and PHQ-9 [67] to track changes in mental health symptoms over time. Additionally, a mixed-methods approach incorporating passive data collection such as longitudinal digital behavioral tracking could offer deeper insights into long-term impact. This may include analyzing user engagement patterns, participation in follow-up sessions, and adherence to ongoing support [68]. Furthermore, metrics such as session frequency, interaction sentiment analysis, and continued chatbot usage could serve as indirect indicators of long-term well-being improvements [69].

The study used English as the language of interaction. We acknowledge that this excluded a portion of the Afghan population who are not proficient in English. Given that GPT-4’s proficiency is higher in English [70] compared to local Afghan languages such as Dari (Persian) and Pashto, the choice to use English was made to ensure better performance and engagement. However, this limitation may have influenced the depth of participant self-expression, as individuals often communicate emotions more naturally in their native language [71]. The challenge of expressing emotions in English could have impacted the quality of interactions, as the chatbot may have struggled to fully comprehend the nuanced emotions being expressed, limiting its ability to provide contextually appropriate support. This limitation highlights the need for future studies to explore local language adaptations. Developing AI tools tailored to local languages, or incorporating multilingual models, would allow for broader inclusivity and greater accessibility for non-English-speaking Afghan women. Other cultural adaptations, such as incorporating regional dialects and culturally specific expressions [72], could foster greater trust and connection with users while respecting their cultural context. Future studies could also examine the feasibility and challenges of implementing AI-driven mental health interventions in local languages, as well as explore how language adaptation might impact the effectiveness and engagement of such interventions. By incorporating these elements, the study's inclusivity and applicability would be significantly enhanced, ensuring that interventions are more accessible to non-English-speaking populations and more attuned to their cultural contexts.

The scope of interaction with the AI was focused primarily on sharing problems with a focus on discussing potential solutions. Future interventions could explore incorporating Cognitive Behavioral Therapy (CBT) or other therapeutic techniques to enhance the breadth and depth of intervention strategies.

Our agents did not have interaction memory within each session, meaning they did not retain previous messages from the participant. Given the highly sensitive nature of the target population and the stigma surrounding mental health in Afghanistan, memory features were excluded to ensure participants' comfort, as the chatbot's ability to recall past conversations could have potentially caused discomfort or unease. Enabling memory features in future implementations presents both potential benefits and risks [73]. On the one hand, incorporating memory could enhance personalization and continuity, allowing the chatbot to recall past interactions, and provide more tailored responses. This could foster a stronger sense of support and improve user adherence and engagement by creating a more natural and cohesive conversational flow [74, 75]. On the other hand, the use of memory raises important ethical and privacy considerations. Participants may feel uneasy knowing that past conversations are being stored, which could lead to self-censorship or reduced openness in discussing sensitive topics [74]. We suggest future research should explore secure, transparent and user-controlled memory features, allowing participants to manage stored interactions while maintaining data privacy [73]. Examining participant preferences regarding memory retention and evaluating its impact on therapeutic outcomes could provide valuable insights into the trade-offs between continuity and psychological safety. Future research may also consider incorporating interaction memory for a subset of participants to yield insights into how continuity influences therapeutic outcomes.

Caution should be exercised when applying AI for therapeutic purposes, especially in studies targeting vulnerable populations. Participants in our study were explicitly instructed not to disclose sensitive information and were asked to confirm they were in a safe environment before starting the chat with the chatbot. While AI models like GPT have inherent rules and safeguards, additional stringent guidelines should be implemented to ensure ethical conduct and participant safety. Beyond these precautions, the ethical implications of AI-driven mental health interventions warrant deeper consideration [75]. Data privacy remains a critical concern, as AI-based systems process sensitive personal information [73]. Ensuring secure data storage, anonymization, and strict access controls is essential to protect participants' confidentiality [32, 76]. Additionally, informed consent should be comprehensive, clearly communicating how data is used, stored, and protected, as well as the limitations of AI-based support [77].

Another ethical challenge is the potential dependency on AI-driven mental health interventions. While these tools can offer scalable and accessible support, they should complement, rather than replace, traditional mental health care [32]. Overreliance on AI could deter individuals from seeking professional help when needed [75]. Moreover, AI lacks real-time intervention capabilities; unlike human therapists, it cannot physically intervene or assess a user's environment in crisis situations [75]. This limitation underscores the importance of integrating AI systems with human oversight, particularly in professional settings, to ensure timely and appropriate intervention when necessary. Future research should explore strategies to integrate AI responsibly, ensuring that users are encouraged to seek appropriate clinical support when needed.

While our study examined linguistic analysis without direct instructions on language use to GPT as personal pronouns, tones, emotions, and LSM in the context of AI therapy, future studies could consider instructing AI systems to emphasize positive emotions, enhance language style matching with participants, or integrate specific therapeutic techniques tailored to individual needs. These limitations highlight the preliminary nature of our findings and underscore the need for larger-scale studies with diverse populations and extended intervention durations to further elucidate the potential of AI-driven interventions in mental health support. Future research should explore refining AI algorithms to better align with cultural nuances and specific mental health needs of diverse populations.

## VI. CONCLUSION

This study investigated the effectiveness of utilizing GPT-4 as a supportive listener in mental health interventions for Afghan women facing diverse challenges. Our findings indicate significant reductions in anxiety and depression scores among participants engaged with the supportive listener compared to control groups. The study underscores the potential of AI-driven interventions, particularly large language models like

GPT-4, to provide accessible and culturally sensitive mental health support. Despite some inherent limitations, such as small sample size and short intervention duration, this study has added to the global literature on positive effects of AI-assisted therapy on improving mental wellbeing of vulnerable women. Given that GPT-4 has demonstrated acceptability and effectiveness, they hold significant potential to become scalable interventions, particularly among hard-to-research populations, to enhance their mental health outcomes.

## ACKNOWLEDGMENT

We would like to acknowledge Tianchen Wang, a master's student at Kyoto University in 2023, for his contribution in implementing the chatbot system during his part-time engagement with our project.